\documentclass[fleqn,usenatbib]{mnras}

\usepackage{newtxtext,newtxmath}

\usepackage[T1]{fontenc}

\DeclareRobustCommand{\VAN}[3]{#2}
\let\VANthebibliography\thebibliography
\def\thebibliography{\DeclareRobustCommand{\VAN}[3]{##3}\VANthebibliography}

\usepackage{graphicx}	
\usepackage{amsmath}	
\usepackage{array}

\usepackage{hyperref}
\usepackage{comment}
\usepackage{newtxtext,newtxmath}
\usepackage{caption}
\usepackage{orcidlink}
\usepackage{xspace}

\newcommand\edtai[1]{{\color{black}#1}}
\newcommand\presub[1]{{\color{black}#1}}   
\newcommand\referee[1]{{\color{black}#1}}   

\newcommand{\ixpe}{\textit{IXPE}\xspace} 
\newcommand {\flux}{erg~s$^{-1}$~cm$^{-2}$}
\newcommand {\lum}{erg~s$^{-1}$}

\title[X-ray polarisation of IC 4329A]{The X-ray polarisation of the Seyfert 1 galaxy IC 4329A}

\author[IXPE Collaboration]{
A. Ingram,$^{1}$\thanks{E-mail: adam.ingram@newcastle.ac.uk}\orcidlink{0000-0002-5311-9078}
M. Ewing,$^{1}$\orcidlink{0000-0001-9349-8271}
A. Marinucci,$^{2}$\orcidlink{0000-0002-2055-4946}
D. Tagliacozzo,$^{3}$\orcidlink{0000-0003-3745-0112}
D. J. Rosario,$^{1}$\orcidlink{0000-0002-0001-3587}
A. Veledina,$^{4,5}$\orcidlink{0000-0002-5767-7253}
\newauthor \;
D. E. Kim,$^{6,7,8}$\orcidlink{0000-0001-5717-3736}
F. Marin,$^{9}$\orcidlink{0000-0003-4952-0835}
S. Bianchi,$^{3}$\orcidlink{0000-0002-4622-4240}
J. Poutanen,$^{4}$\orcidlink{0000-0002-0983-0049}
% ...
\\
\\
     Affiliations are shown at the end of the paper 
}

\date{Accepted XXX. Received YYY; in original form ZZZ}

\pubyear{2022}

\begin{document}
\label{firstpage}
\pagerange{\pageref{firstpage}--\pageref{lastpage}}
\maketitle

\begin{abstract}
We present an X-ray spectro-polarimetric analysis of the bright Seyfert galaxy IC 4329A. The \textit{Imaging X-ray Polarimetry Explorer} (\textit{IXPE}) observed the source for $\sim 500$ ks, supported by \textit{XMM-Newton} ($\sim 60$ ks) and \textit{NuSTAR} ($\sim 80$ ks) exposures. We detect polarisation in the 2--8 keV band with $2.97\sigma$ confidence. We report a polarisation degree of 
\edtai{$3.3 \pm 1.1$}
per cent and a polarisation angle of
\edtai{$78\degr \pm 10\degr$ (errors are $1\sigma$ confidence)}.
The X-ray polarisation is consistent with being aligned with the radio jet, albeit partially due to large uncertainties on the radio position angle. We jointly fit the spectra from the three observatories to constrain the presence of a relativistic reflection component. From this, we obtain constraints on the inclination angle to the inner disc
($<39\degr$ at $99$ per cent confidence)
and the disc inner radius
($<11$ gravitational radii at $99$ per cent confidence),
although we note that modelling systematics in practice add to the quoted statistical error. Our spectro-polarimetric modelling indicates that the 2--8 keV polarisation is consistent with being dominated by emission directly observed from the X-ray corona, but the polarisation of the reflection component is completely unconstrained. Our
constraints on viewer inclination and polarisation degree tentatively favour more asymmetric, possibly out-flowing, coronal geometries that produce more highly polarised emission, but the coronal geometry is unconstrained at the $3\sigma$ level.
\end{abstract}




\section{Introduction}

Active galactic nuclei (AGNs) are powered by accretion of material from the host galaxy onto the central supermassive black hole. An optically thick, geometrically thin disc that radiates a multi-temperature blackbody spectrum peaking in the UV is thought to be present \citep{Shakura1973,Novikov1973}, whereas the X-ray spectrum is typically dominated by a power law with high energy cut-off. The power-law component is thought to result from Compton up-scattering of disc seed photons by a hot thermal population of electrons in a cloud of moderate optical depth ($\tau \sim 0.5-3$) located close to the hole \citep{Thorne1975,Sunyaev1979}, commonly referred to as the \textit{corona}. The geometry and physical nature of the corona are still debated (e.g. \citealt{Poutanen2018}). Proposed geometries include a patchy layer sandwiching the disc \citep{Galeev1979,Haardt94,Stern1995}, a large scale-height accretion flow that forms inside of a truncated disc (the truncated disc model; e.g. \citealt{Eardley1975,Veledina2011a}), and a compact or vertically extended region at the base of the outflowing jet \citep{Martocchia1996,henri1997,miniutti2003,Markoff2005}. Essentially all of these models can reproduce the observed X-ray spectrum, and so a new diagnostic is required to break the impasse.

Modelling of the \textit{reflection} spectrum that results from coronal X-rays irradiating the disc and being re-processed has provided hints. The emergent spectrum includes a Compton hump at $\sim$ 20--30\,keV and an iron emission line at $\sim 6.4$ keV \citep{Matt1991} that is broadened and skewed by disc orbital motion and gravitational redshift \citep{Fabian1989}. The line profile depends on the disc inclination and inner radius as well as the radial dependence of illuminating intensity (the \textit{emissivity profile}), which in turn depends on the geometry of the corona \edtai{(e.g. size and radial/vertical extent)}. Studies that parameterise the emissivity profile as a broken power law often infer very centrally peaked irradiation (\edtai{emissivity of $\epsilon(r) \propto r^{-q}$, where $q\gtrsim 7$}), implying that the corona is very compact (e.g., \citealt{Reynolds2021}). However, considering for example a radially stratified corona approximated by two regions each with their own power-law index and reflection spectrum can produce fits as good as those considering a single compact corona \citep{Basak2017}.

Additionally including timing information can in principle break degeneracies. Features consistent with reverberation lags -- caused by reflected photons taking a longer path than directly observed photons -- have been detected for $\sim$25 AGNs \citep{Kara2016}. Modelling of these reverberation features provides a means to measure black hole mass, but this measurement is likely dependent on the unknown coronal shape \citep{Mastroserio2020}.

\begin{figure*}
\centering
\includegraphics[width=2\columnwidth,trim=0.0cm 0.0cm 8.0cm 14.0cm,clip=true]{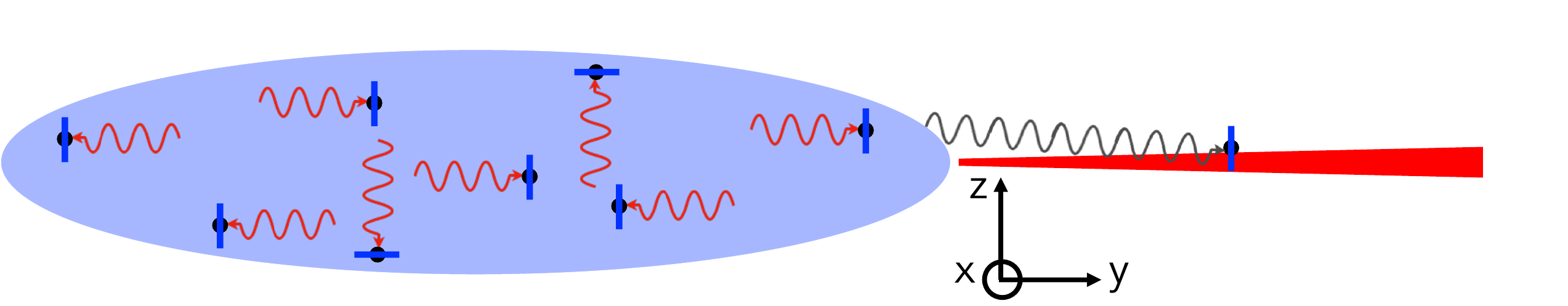}
\vspace{0mm}
\caption{\referee{Schematic diagram demonstrating the basic physics governing X-ray polarisation. The adopted coordinate system is shown, with the observer's sight line being placed in the positive x-direction. The corona and disk are represented respectively by the blue and red regions. The electric field oscillation corresponding to photons before and after a Compton scattering event with an electron (black circles) are colour-coded red and blue, respectively. The trajectory of a photon after a scattering is most likely to be in the plane perpendicular to its electric polarisation vector. Therefore photons that reach the observer vertically polarised (i.e. in the z-direction) were most likely travelling in the x-y plane just before their final scattering, whereas photons that reach the observer horizontally polarised (i.e. in the y-direction) were most likely travelling in the x-z plane just before their final scattering. This means that a corona extended in the x-y plane will overall be vertically polarised because more photons were travelling in the x-y plane before their final scattering due to the higher scattering optical depth in this plane. Similarly, photons from the corona that scatter off free electrons in the disc (black line) become polarised perpendicular to their initial direction of travel (ignoring relativistic effects). This results in the Compton hump being approximately polarised in the direction of the disc normal.}}
\label{fig:schem}
\end{figure*}

The {\it Imaging X-ray Polarimetry Explorer} (\ixpe), which was launched on 2021 December 9, provides the hitherto inaccessible diagnostic of X-ray polarisation. \referee{The polarisation of the corona, which is determined by the physics of Compton scattering, informs on its degree of asymmetry and its orientation on the sky (see Fig. \ref{fig:schem}). The more asymmetric the projection of the corona on the sky is, the more polarised we measure it to be \citep[e.g.][]{Schnittman2010}. Therefore, for example, the polarisation of a torus-shaped corona is greater for a smaller aspect ratio and a more edge-on viewing angle. The polarisation is expected to align with the minor axis of the corona, such that a horizontally extended corona will be vertically polarised. As for reflection, the iron line should only be very weakly polarised\footnote{\edtai{Fluorescence photons are initially unpolarised, but can pick up a small degree of polarisation via scatterings as they propagate out of the disc atmosphere.}} whereas the reflection continuum (i.e. everything except for the fluorescence lines) can be highly polarised \citep[e.g.][]{Matt1993refl,Poutanen1996refl,Dovciak2011}.}

\ixpe is sensitive to 2--8 keV photons, and has so far observed four radio-quiet AGNs: the Circinus galaxy \citep{Ursini2023}, MCG-05-23-16 \citep{Marinucci2022}, NGC 4151 \citep{Gianolli2023} and, most recently IC 4329A. IC 4329A \edtai{(redshift $z=0.0161$, \citealt{Willmer1991})} is a Seyfert Type 1.2 galaxy \citep{Veron-Cetty2006}\edtai{, meaning that its optical spectrum includes both broad and narrow emission lines and its X-ray spectrum is relatively un-obscured}. This indicates in the unification scheme \citep{Urry1995} that the inclination angle is small enough for our sight line to the supermassive black hole to not be blocked by the dusty torus\edtai{, such that we see X-rays directly from the accretion flow and broad optical lines from gas rapidly orbiting the black hole (the broad line region)}. A recent optical reverberation campaign \citep{Bentz2023} further backed up this hypothesis and provided a black hole mass measurement of $M \approx 7 \times 10^7$M$_\odot$, broadly consistent with earlier estimates \citep[e.g.][]{Gonzalez-Martin2012}. An X-ray reverberation feature consistent with this mass has been observed from the source \citep{Kara2016}. IC 4329A was selected for an \ixpe observation primarily because it is one of the brightest AGN in the X-ray sky, with a 2--10 keV flux in the range $F_{2-10} \sim (0.1-1.8) \times 10^{-10}$\flux\  \citep[e.g.,][]{Beckmann2006}.
Optical images show that we view the host galaxy edge-on, with the dust lane associated with the galaxy's spiral arms cutting across our view of the nucleus \citep{Martin1982,Mehdipour2018,Bentz2023}. This implies that the AGN torus is misaligned with the galaxy rotation axis, which could be due to a past interaction with the nearby giant lenticular galaxy IC 4329; although we note that such a misalignment is apparently fairly common \citep[e.g.][]{Schmitt1997,Nagar1999,Middleton2016}.

Here, we present the results of an observing campaign on IC 4329A utilising \ixpe as well as the \textit{X-ray Multi-Mirror Mission} (\textit{XMM-Newton}) and the \textit{Nuclear Spectroscopic Telescope ARray} (\textit{NuSTAR}). In Section~\ref{sec:obs}, we detail our data reduction procedure. In Section~\ref{sec:results} we present our polarimetric analysis as well as spectral and spectro-polarimetric fits. In Section~\ref{sec:discussion} we discuss our results in the context of Comptonization models of polarisation and multi-wavelength observations of the system, and we summarise our results in Section~\ref{sec:conclusions}.

\begin{figure}
\centering
\includegraphics[width=\columnwidth,trim=2.0cm 1.4cm 2.0cm 5.0cm,clip=true]{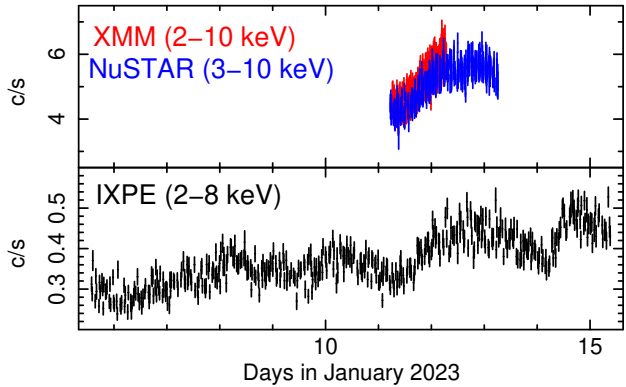}
\vspace{-5mm}
\caption{\referee{Light curves (counts per second) of IC 4329A as observed by \textit{XMM-Newton} EPIC-pn (2--10 keV, 100 s time bins, red), \textit{NuSTAR} (3--10 keV, 200 s time bins, blue) and \ixpe (2--8 keV, 1 ks time bins black). For \textit{NuSTAR} and \ixpe, separate detectors have been summed. Time is represented as days in January 2023, such that 10 is midnight on January 10.}}
\label{fig:lc}
\end{figure}

\section{Observations and data reduction}
\label{sec:obs}

\ixpe observed IC 4329A on 2023 January 5--15  (OBSID 01003601) for a useful exposure time of 458~ks. During this observation, \textit{XMM-Newton} also observed the source on January 11--12  (OBSID 0890670201) for an exposure time of 62~ks, and \textit{NuSTAR} observed it on January 11--13  (OBSID 60701015002) for an exposure of 82~ks. \referee{The light curves of all three of these observations are presented in Fig. \ref{fig:lc}.} The following sub-sections detail the data reduction procedure we employed for the three observatories. Throughout, we make use of \texttt{ftools} from the \textsc{heasoft} package (v6.31.1).

\subsection{\ixpe}

We downloaded cleaned level 2 event files, taken by the three Gas Pixel Detectors \citep{Costa2001}, from the \edtai{High Energy Astrophysics Science Archive Research Center},\footnote{\url{https://heasarc.gsfc.nasa.gov/docs/ixpe/analysis/IXPE-SOC-DOC-009-UserGuide-Software.pdf}} that had been calibrated with the standard pipeline. We use the public tool \textsc{ixpeobssim} \citep[v30.0.0;][]{Baldini2022} to extract Stokes parameters individually for the three detector units (DUs), employing the most recent \ixpe calibration database files (CALDB 20221020). We employ a circular source region centered on the source with radius 60 arcsec. The background region is an annulus, also centered on the source, with inner and outer radii of 150 and 300 arcsec, respectively. \edtai{These regions are chosen such that the source region captures maximal source photons and the background region is as large as possible whilst avoiding source contamination.}

We employ two methods to determine the polarisation in the 2--8 keV energy band. First is the \texttt{pcube} algorithm, which sums Stokes parameters of individual events within the extraction region and specified energy range.
The polarisation degree (PD) relates to the summed Stokes parameters $I$, $Q$ and $U$ as $p = \sqrt{(Q/I)^2+(U/I)^2}$, and the polarisation angle (PA) is $\psi = (1/2) \arctan(U/Q)$ \edtai{(i.e. in this notation, $Q$ and $U$ are corrected for the modulation factor)}. \edtai{Uncertainties} are calculated using standard formulae \citep{Muleri2022}. We extract summed Stokes parameters separately from the source and background regions for each DU. Taking advantage of the linear properties of Stokes parameters, we combine DUs by summing Stokes parameters and account for background by subtracting the background region Stokes parameters from the source region Stokes parameters. We combine \edtai{uncertainties} by summing in quadrature.

The second method is to extract $I$, $Q$ and $U$ source and background spectra for each DU using the \texttt{pha} algorithm. We group the $Q$ and $U$ spectra into energy bins $0.2$ keV wide, and for $I$ instead employ the native energy bins of width $0.04$ keV. We extract $I$, $Q$ and $U$ spectra employing both `weighted' and `unweighted' methods. A weighted analysis gives extra weight to events with better characterised photo-electron tracks \citep{DiMarco2022}, whereas this extra information is ignored in an unweighted analysis. Finally, we also extract weighted $I$, $Q$ and $U$ spectra from event files filtered using the background rejection method of \cite{DiMarco2023}.

Using \textsc{xspec} v12.12.1 \citep{Arnaud1996}, \edtai{we fit the model}
\begin{equation}
   \texttt{TBabs} \times \texttt{const} \times \texttt{polconst} \times \texttt{po},
   \label{eqn:simple}
\end{equation}
\edtai{simultaneously to $I$, $Q$ and $U$ spectra.}
\edtai{Here,} \texttt{TBabs} \citep{Wilms2000} represents line of sight absorption, for which we fix the hydrogen column density to $N_{\rm H} = 4.1\times 10^{20}~{\rm cm^{-2}}$ \citep{HI4PI2016}. The constant \texttt{const} accounts for cross-calibration between different DUs. It is therefore tied between $I$, $Q$ and $U$ of the same DU, but allowed to differ between DUs. \texttt{po} is simply a power law (specific photon flux $\propto E^{-\Gamma}$). \texttt{polconst} imparts a constant (i.e. independent of energy) PD and PA to the model it operates on. We fit the above model simultaneously to all \ixpe DUs, leaving free the power-law index and normalisation, the calibration constants, and the PD and PA. \edtai{Although we consider the three DUs separately in the fits, we combine them in plots for clarity.}

\subsection{\textit{XMM-Newton}}

\textit{XMM-Newton} observed the source with the EPIC CCD cameras: the pn \citep{struder01} and the two MOS \citep{turner01}, operated in small window and medium filter mode.
Data from the MOS detectors are not included in our analysis due to pile-up. The data from the pn camera show no significant pile-up as indicated by the \texttt{epatplot} output. The extraction radii and the optimal time cuts for flaring particle background were computed with SAS 20 \citep{gabr04}, via an iterative process which leads to the maximization of the signal-to-noise ratio (SNR) in the 0.5--10~keV energy band, similar to the approach described in \citet{pico04}. The resulting optimal extraction radii for the source and the background spectra are 40 and 50~arcsec, respectively. The net exposure time for the pn time-averaged spectrum is 61.7~ks. The keyword \texttt{applyabsfluxcorr=yes} was applied in the \texttt{arfgen} task to correct the pn effective area, removing residuals between simultaneous \textit{XMM-Newton} pn and \textit{NuSTAR} spectra.

\subsection{\textit{NuSTAR}}

\textit{NuSTAR} \citep{nustar} observed the source with its two co-aligned X-ray telescopes with corresponding Focal Plane Module A (FPMA) and B (FPMB). The Level 1 data products were processed with the \textit{NuSTAR} Data Analysis Software (NuSTARDAS) package (v. 2.1.2). Cleaned event files (level 2 data products) were produced and calibrated using standard filtering criteria with the \texttt{nupipeline} task and the calibration files available in the \textit{NuSTAR} calibration database (CALDB 20220510). Extraction radii for the source and background spectra were 40 and 60 arcsec, FPMA spectra were binned in order not to over-sample the instrumental resolution more than a factor of 2.5 and to have a SNR  greater than 5 in each spectral channel, the same energy binning was then applied to the FPMB spectra. The net observing times for the FPMA and the FPMB data sets are 82.4 and 81.5~ks, respectively. FPMA and FPMB spectra are considered separately for the purposes of spectral fitting, but are combined in plots for clarity.

\section{Results}
\label{sec:results}

\subsection{X-ray polarisation}

\begin{figure}
\centering
\includegraphics[width=\columnwidth,trim=2.0cm 1.4cm 2.0cm 1.4cm,clip=true]{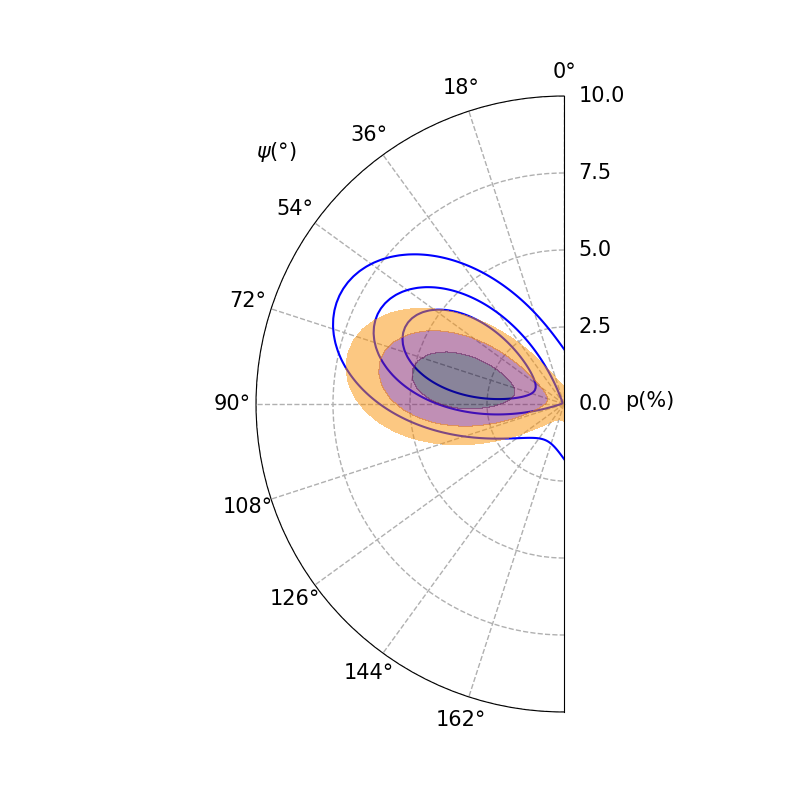}
\vspace{-5mm}
\caption{Constraints on the PD and PA \edtai{(measured east of north)} in the 2--8 keV band. 
Contours at 68, 90 and 99 per cent confidence levels are shown.
Blue lines result from an unweighted analysis with the \texttt{pcube} algorithm. Block colours result from a weighted spectro-polarimetric analysis within \textsc{xspec} that additionally makes use of the new background rejection method. We see that the weighted analysis provides consistent but more tightly constrained results.}
\label{fig:pcube}
\end{figure}

We first measure the 2--8 keV polarisation using the model-independent \texttt{pcube} algorithm on unweighted data. \edtai{We find $p=3.4\pm1.6$ per cent and $\psi=64\degr\pm13\degr$ ($1\sigma$ uncertainties) for the PD and PA, and we plot statistical confidence contours}
in Fig. \ref{fig:pcube} (blue lines). 

Currently, \texttt{pcube} does not feature the functionality to improve signal to noise by including event weights. Therefore to make use of a weighted analysis, we employ the alternative method of fitting weighted $I$, $Q$ and $U$ spectra (produced utilising the background rejection method) with a simple spectro-polarimetric model \edtai{(Equation \ref{eqn:simple})} within \textsc{xspec}. The best-fitting reduced $\chi^2$ is $\chi^2/{\rm d.o.f.} = 659 / 614$. \edtai{This is an acceptable fit, although we note that a more complex model is required when we also include \textit{XMM-Newton} and \textit{NuSTAR} data (see the following two sections)}.

\edtai{We find $p = 3.3 \pm 1.1$ per cent and $\psi = 78\degr \pm 10\degr$ ($1\sigma$ errors for a single parameter of interest), and plot the confidence contours}
as block colours in Fig. \ref{fig:pcube}. We see that, as expected, the results are consistent with the \texttt{pcube} method but with smaller uncertainties. The $99$ per cent contour (representing $3\sigma$ for a two dimensional distribution) still does not quite close, meaning that we fall just short of a $3\sigma$ detection of polarisation. The formal confidence is $98.78$ per cent, or $2.97\sigma$. 
The $3\sigma$ upper limit on \edtai{the} PD is $6.2$ per cent.

As a check, we also
\edtai{performed}
the fitting technique on unweighted data extracted without the background rejection method \citep{DiMarco2023} and find results very close \edtai{($p=3.7 \pm 1.2$ per cent, $\psi = 71\degr\pm10\degr$)} to those yielded by \texttt{pcube}. \edtai{We also find that using or not using the background rejection method has no effect on the results within our chosen rounding convention.}

\referee{An alternative way to assess the significance of a polarisation measurement is to compare the measured PD to the minimum detectable polarisation (MDP). This is the statistical upper limit on polarisation that we would measure for an unpolarised source. Following convention \citep[e.g.][]{Muleri2022}, we consider MDP$_{99}$, whereby the statistical upper limit is $99$ per cent confidence. We calculate MDP$_{99} = 4.5$ per cent for the unweighted analysis and MDP$_{99} = 4.0$ for the weighted analysis. The measured polarisation is thus lower than the MDP, meaning that the confidence with which we can rule out the source being unpolarised is less than $99$ per cent ($3~\sigma$). This is consistent with the $99$ per cent confidence contours not fully closing in Fig. \ref{fig:pcube}.}


\subsection{X-ray spectrum}

We jointly fit the \ixpe, \textit{NuSTAR} and {\it XMM-Newton} Stokes $I$ spectra. We consider the usual energy ranges of 2--8 keV and 3--79 keV for \ixpe and \textit{NuSTAR}, respectively. For {\it XMM-Newton}, we consider the energy range 2--10 keV, leaving the more involved treatment of ionised absorbers necessary to reproduce the soft X-rays \citep{Mehdipour2018} to future work. We employ the following model, referred to hereafter as Model 1:
\begin{equation} \label{eqn:specmod} 
 \texttt{mbpo}\times\texttt{zTBabs}\times\texttt{TBabs}\times [ \texttt{nthComp} + \texttt{relxillCp} + \texttt{xillverCp} ].
\end{equation}
Here, \texttt{nthcomp} \citep{Zdziarski1996,Zycki1999} represents radiation produced by thermal Compton up-scattering in the corona, \texttt{relxillCp} \citep{Garcia2014} represents relativistic reflection from the accretion disc, and \texttt{xillverCp} \citep{Garcia2013} represents non-relativistic reflection from the distant molecular torus. \texttt{tbabs} \citep{Wilms2000} represents line of sight absorption within our Galaxy (we continue to fix $N_{\rm H} = 4.1\times 10^{20}~{\rm cm^{-2}}$), whereas \texttt{ztbabs} represents absorption within the host galaxy, perhaps including material local to the AGN. For this we leave the column density as a free parameter. Finally, \texttt{mbpo} \citep{Ingram2017} is required to circumvent cross-calibration discrepancies between the three observatories. It multiplies the total model by a broken power law with index $\Delta \Gamma_1$ for $E<E_{\rm br}$ and $\Delta \Gamma_2$ for $E>E_{\rm br}$. A single corrective power law (e.g. $\Delta\Gamma_1=\Delta\Gamma_2$) has previously been required to jointly model {\it XMM-Newton} and \textit{NuSTAR} data \citep{Ingram2017}, whereas a broken power law has been required to obtain agreement between \ixpe and other observatories such as {\it NICER} and \textit{NuSTAR} \citep{Krawczynski2022}. \edtai{For $\Delta\Gamma_1=\Delta\Gamma_2=0$, the model simply reduces to a multiplicative calibration constant.}
\referee{Note that dealing with cross-calibration in this way adjusts the \textit{model} and not the data, such that a common set of model parameters can be used to describe the source spectrum but discrepancies between observatories are preserved in plots.}

\begin{figure}
\centering
\includegraphics[width=\columnwidth,trim=1.2cm 0.0cm 2.5cm 0.0cm,clip=true]{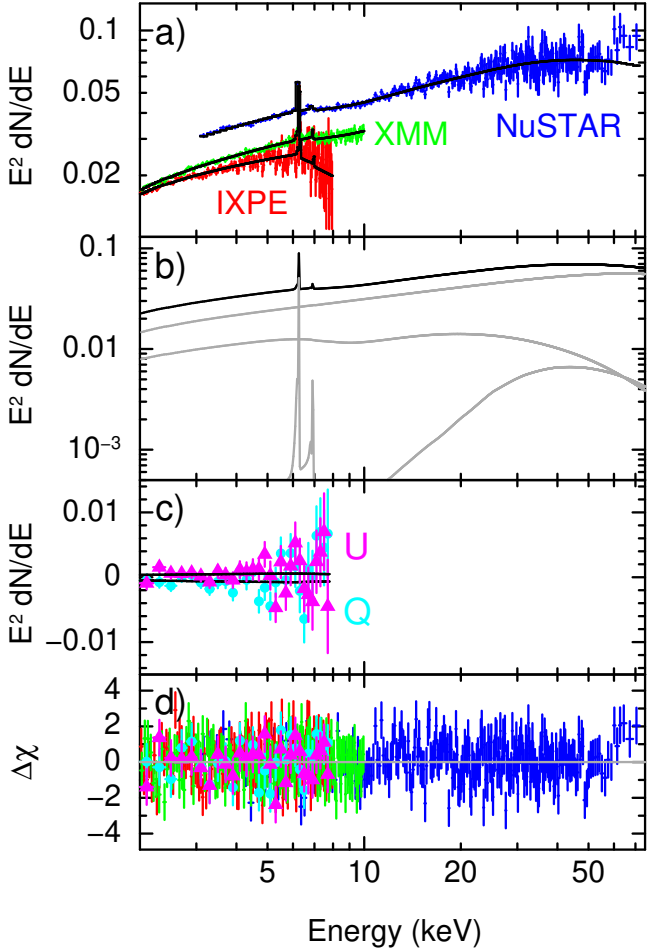}
\vspace{-8mm}
\caption{Spectral energy distribution for Model 1 (Equation \ref{eqn:specmod}). \textit{(a)} Best-fitting spectral model (black) with unfolded data. \textit{(b)} Best-fitting model only (black) with individual components (grey). From top to bottom these are \texttt{nthcomp}, \texttt{relxillCp} and \texttt{xillverCp}. \textit{(c)} Unfolded Stokes $Q$ (cyan circles) and $U$ (magenta triangles) observed by \ixpe with best-fitting model (black lines, top is $U$ and bottom is $Q$). \textit{(d)} Contributions to fit statistic $\chi$.
$dN/dE$ is in units of photons cm$^{-2}$ s$^{-1}$ keV$^{-1}$.}
\label{fig:eeuf}
\end{figure}

\begin{table}
\caption{The best-fitting Model 1 (Equation \ref{eqn:specmod}) parameters from our joint spectral analysis.}
\begin{center}
\begin{tabular}{ c c c c }
\hline
\multicolumn{4}{c}{\edtai{Model 1}} \\
\multicolumn{4}{c}{\texttt{zTBabs}$\times$\texttt{TBabs}$\times$(\texttt{nthcomp} + \texttt{relxillCp} + \texttt{xillverCp})} \\
\hline
{Component} & {Parameter} & {Units} & {Value} \\
\hline
 \texttt{zTBabs} & $N_\text{H}$ & $10^{22}$ ~cm$^{-2}$  & $0.61_{-0.05}^{+0.06}$ \\
\hline
 \texttt{TBabs} & $N_\text{H}$ & $10^{22}$ ~cm$^{-2}$  & $\equiv 0.041$ \\
\hline
 \texttt{nthcomp} & $\Gamma$ &  & $1.66_{-0.01}^{+0.02}$ \\
                    & $kT_\text{bb}$ & keV  & $\equiv 0.05$ \\
                    & $kT_\text{e}$ & keV  & $39_{-9}^{+11}$ \\
                  & \texttt{norm} & $10^{-4}$ & $4.2_{-0.2}^{+0.3}$ \\
\hline
 \texttt{relxillCp} & $i$ & deg & $3.0_{-3}^{+23}$ \\
                  & $a$           &  & $\equiv 0.998$ \\ 
                  & $r_\text{in}$ & $r_\text{g}$ & $2.1_{-0.7}^{+1.3}$ \\
                  & $\log{\xi}$ &   & $4.4_{-0.11}^{+0.11}$ \\
                  & $A_{\rm Fe}$ & solar & $8.3_{-3.0}^{+1.5}$ \\
                  & $n_{\rm e}$ & $10^{15}~{\rm cm}^{-3}$ & $\equiv 1$ \\    
                  & $\epsilon(r)$ & & $\equiv r^{-3}$ \\                      
                  & $f_\text{refl}$ & & $0.25_{-0.11}^{+0.13}$\\

\hline
 \texttt{xillverCp} & $f_\text{refl}$ & & $0.16_{-0.03}^{+0.05}$ \\
                  & $\Delta z$ & $10^{-3}$ & $7.33_{-1.88}^{+0.94}$\\
                  & $\log\xi$ &  & $\equiv 0$\\
\hline
\multicolumn{4}{c}{Cross-calibration} \\
\hline
 \texttt{mbpo} &$\Delta\Gamma_1$& & $\equiv 0$\\
NuSTAR FPMB & \text{N}& & $1.04_{-0.01}^{+0.01}$\\
\hline
 \texttt{mbpo} & $\Delta\Gamma_1$&$10^{-2}$& $0.30_{-1.2}^{+1.4}$\\
 XMM EPIC-pn & $N$& &$0.75_{-0.01}^{+0.01}$ \\
\hline
 \texttt{mbpo} & $\Delta\Gamma_1$&$10^{-2}$&$-11_{-2}^{+3}$\\
 IXPE DU1& $N$ & &$0.71_{-0.01}^{+0.01}$\\

\hline
 \texttt{mbpo} & $\Delta\Gamma_1$& $10^{-2}$ & $ -9.8_{-2.9}^{+3.4}$\\
 IXPE DU2 &$\Delta\Gamma_2$& &$-4.6_{-4.4}^{+3.1}$\\
                  &$E_{\rm br}$&keV& $6.8_{-0.3}^{+0.5}$\\
                  &$N$& &$0.66_{-0.02}^{+0.02}$\\

\hline
 \texttt{mbpo} & $\Delta\Gamma_1$&$10^{-2}$&$-9.7_{-4.1}^{+3.6}$ \\
 IXPE DU3 & $\Delta\Gamma_2$&&$-1.0_{-1.6}^{+0.7}$ \\
                  &$E_{\rm br}$&keV&$6.0_{-0.8}^{+1.3}$ \\
                  &$N$&  &$0.65_{-0.03}^{+0.02}$\\
\hline
\end{tabular}
\begin{flushleft}{\textit{Note}: Errors are 90 per cent confidence\edtai{, and a $\equiv$ symbol indicates that the parameter is fixed/hardwired}. Reduced $\chi^2$ is $\chi^2/{\rm d.o.f.} = 1093/1047$. A gravitational radius is $r_{\rm g}=GM/c^2$. The 3--79 keV \textit{NuSTAR} FPMA flux is $F_{3-79}=2.7\times10^{-10}$\,\flux, corresponding to isotropic luminosity $L_{3-79} = 1.5 \times 10^{44}$\,\lum\ assuming a Hubble constant of $H_0 = 70$\,km\,s$^{-1}$\,Mpc$^{-1}$. \presub{The 2--8 keV \ixpe DU1 flux is $F_{2-8}=5.3\times10^{-11}$\,\flux.}}\end{flushleft} 
\end{center}
\label{tab:parameters}
\end{table}

We achieve a good fit with
$\chi^2/{\rm d.o.f.} = 1093/1047$. The best-fitting parameters are presented in Table \ref{tab:parameters}. In Fig. \ref{fig:eeuf} we plot the unfolded spectrum (a), the model spectrum with separate components (b) and the contributions to fit statistic $\chi$ (d).
In our best-fitting model, $\sim$66 per cent of the 2--8 keV photons are from the direct component, $\sim$33 per cent are from the relativistic reflector, and $\sim$ 1 per cent are from the distant reflector. All three components are required with high statistical significance. Fitting with only \texttt{nthcomp} yields $\chi^2/{\rm d.o.f.} = 1794/1054$. Including the distant reflector reduces this to
$\chi^2/{\rm d.o.f.} = 1131/1051$, and finally also including the relativistic reflector yields the best fit model ($5.1\sigma$ significance, p-value of $3.7\times 10^{-7}$, according to an F-test). We calculate errors on our best-fitting model parameters by running a Markov Chain Monte Carlo (MCMC) simulation within \textsc{xspec}. We employ the \citet{Goodman2010} algorithm with 256 walkers and a total length of 307,200 steps after a burn-in period of 19,200 steps.


The only free parameters of the \texttt{nthcomp} component are the photon index $\Gamma$, the electron temperature $kT_\text{e}$ and the normalisation \texttt{norm}. For computational convenience, we include \texttt{nthcomp} into our model via \texttt{relxillCp} with the reflection fraction parameter (\texttt{refl\_frac}) set to zero. This returns the illuminating continuum (\texttt{nthcomp}) in exactly the same normalisation as that used for the reflection calculation. This choice therefore makes it straight forward to interpret the reflection fraction parameter of the relativistic and distant reflection components. The seed photon temperature is hardwired to $kT_\text{bb}=0.05$~keV in the \texttt{xillverCp} grid, which is appropriate for an AGN, for which the low energy break is expected to be beyond the low energy bandpass of our observations. Our measured value of $\Gamma$ agrees with previous studies \citep[e.g.][]{Brenneman2014}. Our inferred electron temperature broadly agrees with previous \textit{NuSTAR} and \textit{Suzaku} observations \citep{Brenneman2014,Brenneman2014a}.

\begin{figure}
\centering
\includegraphics[width=\columnwidth,trim=1.0cm 1.0cm 2.5cm 11.0cm,clip=true]{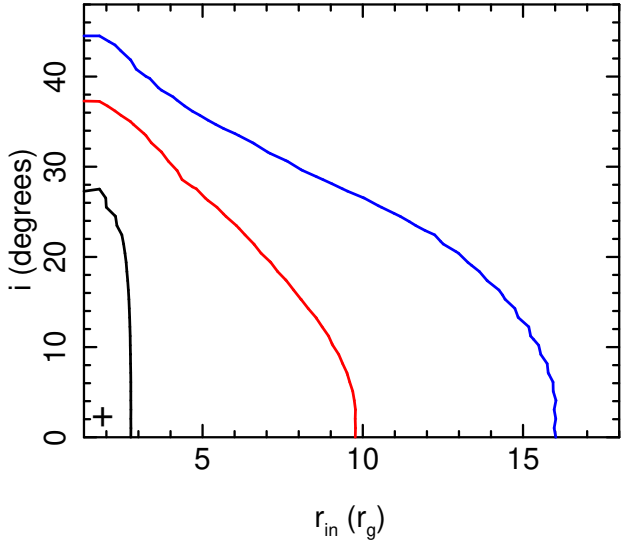}
\vspace{-7mm}
\caption{Constraints on the inclination angle $i$ and the disc inner radius $r_{\rm in}$ based on our Model 1 spectral fit.
The 1, 2, and 3$\sigma$  confidence contours are shown with black, red and blue curves, respectively. 
The best fit is represented by a black cross.}
\label{fig:incrin}
\end{figure}

For the relativistic reflector (\texttt{relxillCp}), we tie $\Gamma$, $kT_{\rm e}$ and \texttt{norm} to their \texttt{nthcomp} values, fix the reflection emissivity profile to $\epsilon(r) \propto r^{-3}$ and freeze the disc density to $n_{\rm e} = 10^{15}~{\rm cm}^{-3}$. We freeze the black hole spin parameter to its maximum value of $a=0.998$ and leave the disc inner radius $r_{\rm in}$ free. This enables us to explore the widest possible range of $r_{\rm in}$ values without it becoming smaller than the innermost stable circular orbit (ISCO) during the fitting procedure. Although the spin itself does influence the model outputs via the photon geodesics, these are very subtle higher order effects. The other free parameters are inclination angle $i$, the logarithm of the ionisation parameter $\log\xi$ (where $\xi$ is in units of erg cm s$^{-1}$, and we take $\log$ to mean log to the base 10 throughout), the iron abundance relative to solar $A_{\rm Fe}$, and the reflection fraction $f_{\rm refl}$. We set the reflection fraction to a negative value within the model, meaning that the code outputs the reflected component only (conversely, a positive value returns the sum of direct and reflected components). Treating direct and reflected components separately in \textsc{xspec} will later allow us to assign different polarisation properties to each of them. In the tables, we convert the reflection fraction back to a positive value. Low inclination and a small disc inner radius are preferred, but the contour plot in Fig. \ref{fig:incrin} shows that a wide range of values are acceptable within $3\sigma$ confidence. Our constraints are consistent with earlier analyses that used the less sophisticated \texttt{diskline} model on {\it ASCA} data. \citet{Nandra1997} found $i<24\degr$ ($1~\sigma$) whilst freezing $r_{\rm in}= 6~r_{\rm g}$, and \citet{Done2000} inferred a mildly truncated disc whilst achieving equally good fits for a range of different inclinations. 
For the distant reflector, we tie $\Gamma$, $kT_{\rm e}$, \texttt{norm} and $A_{\rm Fe}$ to their \texttt{relxillCp} values, freeze $\log\xi=0$ and leave $f_{\rm refl}$ as a free parameter.

The best-fitting values for $\log\xi$ and $A_{\rm Fe}$ are both rather large. Indeed, at $\log\xi \approx 4.4$ we may expect the disc to be over-ionized such that no features are present in the reflection spectrum. However, Fig.~\ref{fig:eeuf}b shows that a broad iron line and Compton hump can still be seen in the spectrum of the relativistic reflector (second grey line from the top), which occurs because the reflection features are boosted by the very high iron abundance.
This combination of parameters is likely the result of the model mimicking some physics that we are not currently including (see Section \ref{sec:discussion} for a detailed discussion on this)

Throughout we fix the redshift to $z=0.016054$ \citep{Willmer1991}, except for that of the {\it XMM-Newton} and \textit{NuSTAR} \texttt{xillverCp} components, which we tie together but leave free in order to circumvent some apparent calibration issues \edtai{(presumably with the gain scale)}. The resulting small shift to the narrow iron line from the known redshift of $\Delta z \approx 7 \times 10^{-3}$ enormously improves the fit ($\Delta \chi^2 = 34$, $5.7\sigma$ improvement according to an F-test), likely indicating a small error in the {\it XMM-Newton} and \textit{NuSTAR} gain scales. {High resolution {\it Chandra} spectroscopy of IC 4329A has previously confirmed that the most prominent narrow iron line is at $\approx 6.3$ keV \citep{McKernan2004}, which is as expected for neutral iron emission ($6.4$ keV in the rest frame) redshifted by the known redshift of the host galaxy. In our fits, the narrow line is instead at $\approx 6.26$ keV. Moreover, when the {\it XMM-Newton} and \textit{NuSTAR} models are allowed to have different redshifts to one another, $\chi^2$ once again improves significantly, with the redshift being greater for \textit{NuSTAR} and both being significantly discrepant with the host galaxy value. No other parameters are affected though.}
A very similar issue is seen in {\it XMM-Newton} and \textit{NuSTAR} data taken as part of \ixpe campaigns on two other Seyfert 1 galaxies: \mbox{MCG-05-23-16} (\citealt{Marinucci2022}; Tagliacozzo et al. in prep) and \mbox{NGC~4151} \citep{Gianolli2023}.

Table \ref{tab:parameters} lists the best-fitting parameters of the cross-calibration model \texttt{mbpo}. We \edtai{arbitrarily} take \textit{NuSTAR} FPMA as the `ground truth' and find that we only need to multiply the FPMB model by a constant to achieve acceptable agreement with the FPMA. We multiply the {\it XMM-Newton} model by a single power law ($\Delta\Gamma_2=\Delta\Gamma_1$) and find that the index $\Delta\Gamma_1$ is small and consistent with zero within
\edtai{$1 \sigma$}
confidence. For \ixpe, we find that only a single corrective power law is required for DU1, whereas broken power-laws are required for DU2 and DU3 (as is also found for Cygnus X-1; \citealt{Krawczynski2022}). This indicates that the effective area of DU2 and DU3 drops off above $E\sim 7$ keV more steeply than the current calibration files suggest \edtai{(and this is currently being investigated by the calibration team)}, although the slope (i.e. $\Delta\Gamma_2$) is poorly constrained by our fits due to the lack of photons with these energies. \edtai{The lack of $E > 7$ keV photons also means that this calibration issue has no influence on the measured polarisation.}

\begin{figure}
\centering
\includegraphics[width=\columnwidth,trim=1.0cm 1.5cm 2.0cm 1.5cm,clip=true]{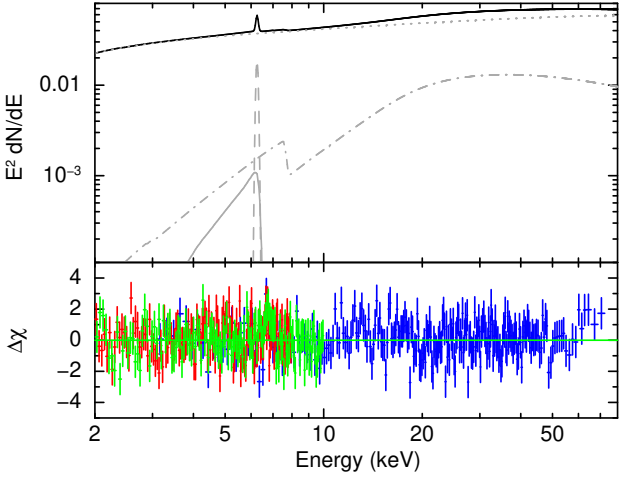}
\vspace{-5mm}
\caption{Best fitting Model 2 spectrum (top) and residuals to \ixpe (red), \textit{NuSTAR} (blue) and \textit{XMM-Newton} (green) data. Different lines represent the total model (black solid), cut off power law (dotted), \texttt{pexriv} (dot-dashed), \texttt{laor} (solid), and the narrow Gaussian (dashed).}
\label{fig:eemo_pexriv}
\end{figure}

\subsection{Spectro-polarimetric fit}

We now include \ixpe Stokes $Q$ and $U$ into our spectral fit. We first extend Model 1 by assigning a constant (i.e. independent of energy) PD and PA to each component with the \texttt{polconst} convolution model.
\referee{This is a very crude approximation for the reflection component, which includes an essentially unpolarised iron line and a potentially highly polarised reflection continuum.}
\referee{Because the iron line is the most prominent contributor to reflection in the \ixpe band,}
we first freeze the PD of both reflection components to zero and leave the \texttt{nthcomp} PD and PA as free parameters. 
\referee{This effectively equates to assuming that the direct coronal emission dominates the 2--8 keV polarisation, which is justifiable as this component contributes the most flux in the band (66 per cent)}. Fitting this model to the spectral-polarimetric data gives
$\chi^2/{\rm d.o.f.} = 1267/1219$. The best-fitting model spectrum and unfolded data for Stokes $Q$ and $U$ are shown in Fig. \ref{fig:eeuf}c. All spectral parameters are almost completely unchanged from the values quoted in Table \ref{tab:parameters}.

\begin{figure}
\centering
\includegraphics[width=\columnwidth,trim=0.0cm 0.0cm 2.0cm 0.0cm,clip=true]{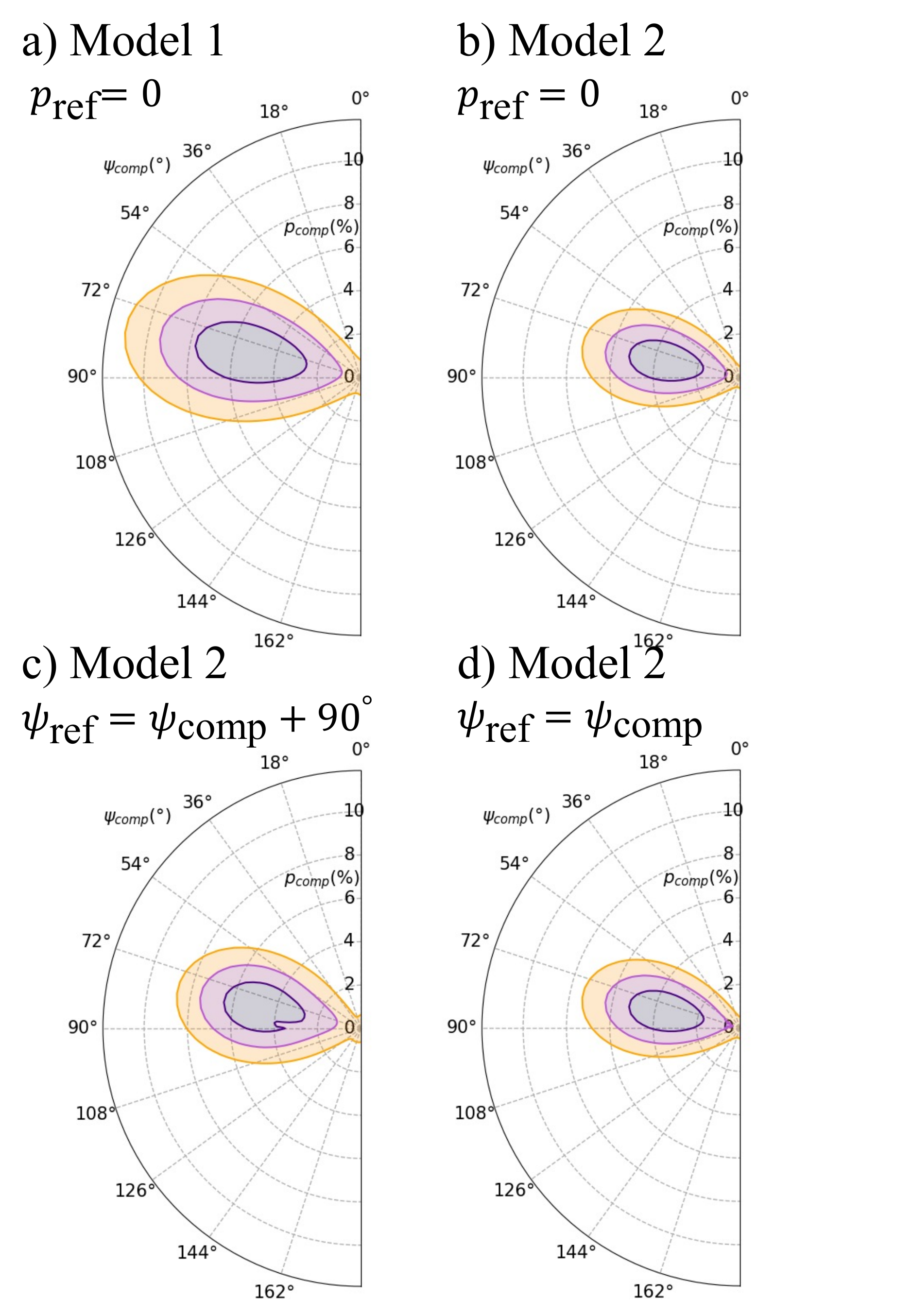}
\vspace{0mm}
\caption{Constraints on the coronal PD ($p_{\rm comp}$) and PA ($\psi_{\rm comp}$) for different modelling assumptions. Three contours correspond to the 68, 90 and 99 per cent confidence levels.   \textit{(a)} Results for Model 1 assuming zero PD of the reflection component $p_{\rm ref} = 0$. \textit{(b)} Results for Model 2, also assuming $p_{\rm ref} = 0$. \textit{(c)} Results for Model 2 with $\psi_{\rm ref} = \psi_{\rm comp} + 90\degr$ and $p_{\rm ref}$ free to vary. \textit{(d)} Results for Model 2 with $\psi_{\rm ref} = \psi_{\rm comp}$ and $p_{\rm ref}$ free to vary. $p_{\rm ref}$ is unconstrained whenever it is left free to vary in the fit.}
\label{fig:ppsicomp}
\end{figure}

\begin{table}
\caption{\edtai{Coronal PD ($p_{\rm comp}$) and PA ($\psi_{\rm comp}$) inferred from our spectro-polrimetric fits for different modelling assumptions.}}
\begin{center}
\begin{tabular}{ l c c }
\hline
{Model} & {$p_{\rm comp}$ (\%)} & {$\psi_{\rm comp}$ (deg)} \\
\hline
a) Model 1, $p_{\rm ref}=0$ & $5.0 \pm 2.6$ & $78 \pm 15$ \\
b) Model 2, $p_{\rm ref}=0$ & $3.4 \pm 1.7 $ & $78 \pm 15$ \\
c) Model 2, $\psi_{\rm ref}=\psi_{\rm comp}+90\degr$ &$4.4 \pm 1.9$ &$78 \pm 14$ \\
c) Model 2, $\psi_{\rm ref}=\psi_{\rm comp}$ & $3.4 \pm 1.7$  &$78 \pm 15$ \\
\hline
\end{tabular}
\begin{flushleft}{\textit{Note}: Errors are $1\sigma$. The confidence contours corresponding to each model are plotted in Fig.~\ref{fig:ppsicomp}.
}\end{flushleft} 
\end{center}
\label{tab:pandpsi}
\end{table}

\begin{table}
\caption{The best-fitting Model 2 (Equation \ref{eqn:specmod2}) parameters from our joint spectral analysis. }
\begin{center}
\begin{tabular}{ c c c c }
\hline
\multicolumn{4}{c}{\edtai{Model 2}} \\
\multicolumn{4}{c}{\texttt{zTBabs}$\times$\texttt{TBabs}$\times$(\texttt{cutoffpl} + \texttt{pexriv} + \texttt{laor} + \texttt{gauss})} \\
\hline
{Component} & {Parameter} & {Units} & {Value} \\
\hline
 \texttt{zTBabs} & $N_{\rm H}$ & $10^{22}~{\rm cm}^{-2}$  & $0.52_{-0.04}^{+0.05}$ \\
\hline
 \texttt{TBabs} & $N_\text{H}$ & $10^{22}$ ~cm$^{-2}$  & $\equiv 0.041$ \\
\hline 
 \texttt{cutoffpl} & $\Gamma$ &  & $1.74_{-0.02}^{+0.02}$ \\
                    & $E_{\rm cut}$ & keV  & $395_{-80}^{+194}$ \\
                  & \texttt{norm} & $10^{-2}$ & $2.14_{0.07}^{+0.07}$ \\
\hline
\texttt{pexriv}  & $f_{\rm refl}$ & & $0.22_{-0.04}^{+0.06}$ \\
                  & $\xi$ & ${\rm erg~cm~s}^{-1}$  & $82_{-62}^{+112}$ \\
\hline
 \texttt{laor}    & $E_{\rm line}$ & keV & $6.26_{-0.02}^{+0.01}$ \\
                    & $i$ & deg & $17_{-15}^{+28}$ \\
                  & $a$ &   & $\equiv 0.998$ \\
                  & $r_{\rm in}$ & $r_{\rm g}$ & $2.8_{-1.4}^{+3.3}$ \\
                  & $\epsilon(r)$ & & $\equiv r^{-3}$ \\         
                  & \texttt{norm} & $10^{-5}$ & $5.0_{-2.4}^{+3.2}$ \\
\hline
 \texttt{gauss} & \texttt{norm} & $10^{-5}$ & $6.38_{-0.86}^{+0.57}$ \\

\hline
\end{tabular}
\begin{flushleft}{\textit{Note}: 
Errors are $90$ per cent confidence.
Reduced $\chi^2$ is $\chi^2/{\rm d.o.f.} = 1110/1047$. The \texttt{mbpo} cross-calibration parameters are consistent with their Model 1 values (Table \ref{tab:parameters}). We allow the iron line centroid energy $E_{\rm line}$ to be a free parameter to circumvent the calibration discrepancies also encountered when fitting Model 1.
}\end{flushleft} 
\end{center}
\label{tab:parameters2}
\end{table}

The resulting \edtai{coronal PD and PA values are quoted in Table \ref{tab:pandpsi} (top row), and their} confidence contours
are plotted in Fig. \ref{fig:ppsicomp}a. We see that the coronal PA here is consistent with the overall 2--8 keV PA presented in Fig.\,\ref{fig:pcube}, whereas the PD is larger, as are the uncertainties. This is because we have assumed the reflection components, which contribute $\sim 34$ per cent of the 2--8 keV flux, to be unpolarised. The resulting dilution increases the best-fitting coronal PD by a factor $\sim 1/0.66$.

\referee{To explore a more realistic scenario with an unpolarised iron line and a polarised reflection continuum,}
we define a new spectral model that we dub Model 2:
\begin{equation}
\texttt{mbpo}\times\texttt{ztbabs}\times\texttt{tbabs}\times [ \texttt{cutoffpl} 
    + \texttt{pexriv} + \texttt{laor} + \texttt{gauss} ].
    \label{eqn:specmod2}
\end{equation}
The direct coronal emission is now represented by an exponentially cut-off power law, \texttt{pexriv} \citep{Magdziarz1995} represents polarised reflection \edtai{(i.e., unlike \texttt{xillverCp}, the \texttt{pexriv} spectrum does not include an iron line)}, \texttt{laor} \citep{Laor1991} is a relativistic iron line and the narrow (width fixed to 0.05 keV) Gaussian is the iron line from the distant reflector. This model is less sophisticated as a spectral model, but enables us to treat the polarisation of the lines separately to other aspects of the reflection signals. When we fit this model to the joint {\it XMM-Newton}, \textit{NuSTAR} and \ixpe Stokes $I$ data, we achieve a fit of similar quality with $\chi^2/{\rm d.o.f.} = 1110/1047$. The best-fitting Model 2 parameters are quoted in Table \ref{tab:parameters2}\edtai{, and the best fitting model is plotted in Fig. \ref{fig:eemo_pexriv}}. Fitting Model 2 without the \texttt{laor} component yields $\chi^2/{\rm d.o.f.} = 1125/1050$, meaning that the relativistically broadened iron line is required with $3.1\sigma$ confidence (according to an F-test). Moreover, the best-fitting $r_{\rm in}$ and $i$ values agree between Model 1 and 2.

\edtai{Although the iron line is consistent between Model 1 and Model 2, the reflection continuum is quite different, with the direct continuum contributing $97$ per cent of the 2--8 keV flux for Model 2 as opposed to $66$ per cent for Model 1. This is because Model 2 makes the crude approximation that the reflection continua of the two reflection components can simply be summed together and represented by one \texttt{pexriv} component. In reality, we expect the relativistic reflector to be highly ionised and the distant reflector to be approximately neutral. This behaviour is captured by Model 1, with the \texttt{relxillCp} spectrum being much softer than the \texttt{xillverCp} spectrum due to its higher ionisation parameter. Model 2 is not as flexible: the fit can either choose a low ionisation to reproduce the distant reflection continuum or a high ionisation to reproduce the relativistic reflection continuum. The best fitting ionisation parameter is low ($\xi \approx 80~{\rm erg}~{\rm cm}~{\rm s}^{-1}$), and thus the \texttt{pexriv} contribution is very small in the soft X-rays. We tried including an extra neutral reflection continuum in Model 2 (\texttt{pexrav}), but this model proved highly degenerate.}

We then include the Stokes $Q$ and $U$ spectra into our Model~2 fit, again by convolving each additive spectral component with \texttt{polconst}. We freeze the polarisation of the two line components  (\texttt{laor} and \texttt{gauss}) to zero and try several assumptions for the PD and PA of the direct and reflected components: $p_{\rm comp}$, $\psi_{\rm comp}$, $p_{\rm ref}$ and $\psi_{\rm ref}$. The resulting confidence contours for 
$p_{\rm comp}$ and $\psi_{\rm comp}$ are plotted in Fig \ref{fig:ppsicomp}b-d\edtai{, and their values are quoted in Table \ref{tab:pandpsi}}. Panel~b shows the results assuming $p_{\rm ref}=0$. We see that the PA is consistent with the Model 1 result (panel~a), but the PD is now smaller. This is because of the much smaller 2--8 keV contribution from reflection in the Model 2 fit (3 per cent as opposed to 34 per cent). The dilution from the unpolarised component is therefore smaller and the coronal PD comes out at a similar value to the overall 2--8 keV PD. Panel~c shows the results assuming direct and reflected components are polarised perpendicular to one another, as would be expected if, for example, the corona is vertically extended \referee{(see Fig. \ref{fig:schem})}. 
In this case, the total polarised flux is the difference between that of the two spectral components. The coronal PD is therefore larger than in panel~b to offset this subtraction. The PD of the reflection component is completely unconstrained (all values from 0 to 100 per cent are acceptable within $1\sigma$ confidence), and the small difference between panels~b and c illustrates that its best-fitting value is small. Panel~d presents the results assuming the two components have parallel polarisation (as would occur if, for example, the corona is radially extended\referee{; see Fig. \ref{fig:schem}}), meaning that the polarised flux is summed over the two components.
The coronal PD is therefore slightly smaller than in panel b, due to a small (albeit still completely unconstrained) contribution from the reflection component.

\section{Discussion}
\label{sec:discussion}

\subsection{Multi-wavelength comparison}

Although \edtai{not significant} above the $3\sigma$ level, we find that the
\edtai{$1 \sigma$}
confidence range on the 2--8  keV PA of IC 4329A is
$\psi = 78\degr\pm10\degr$.
Our spectro-polarimetric analysis yields similar constraints on the PA of the corona. Optical and infrared polarisation \citep{Wolstencroft1995} instead align with the $\approx 45\degr$ position angle of the galaxy disc (see, e.g., Fig.~1 of \citealt{Bentz2023}), indicating that it is consistent with dust scattering in the galactic dust lane. Radio emission can be seen to extend along the position angle $\sim$90\degr\ in the highest resolution 1.5\,GHz and 5\,GHz {Very Large Array} ({VLA}) images in the literature \citep[][beam diameters $\sim 3\arcsec$ and $\sim 1.5\arcsec$, respectively]{Unger1987}. These images resemble a marginally resolved jet, with extended emission appearing only to the West of the core.

\begin{figure}
\centering
\includegraphics[width=\columnwidth,trim=0.0cm 0.0cm 1.0cm 0.5cm,clip=true]{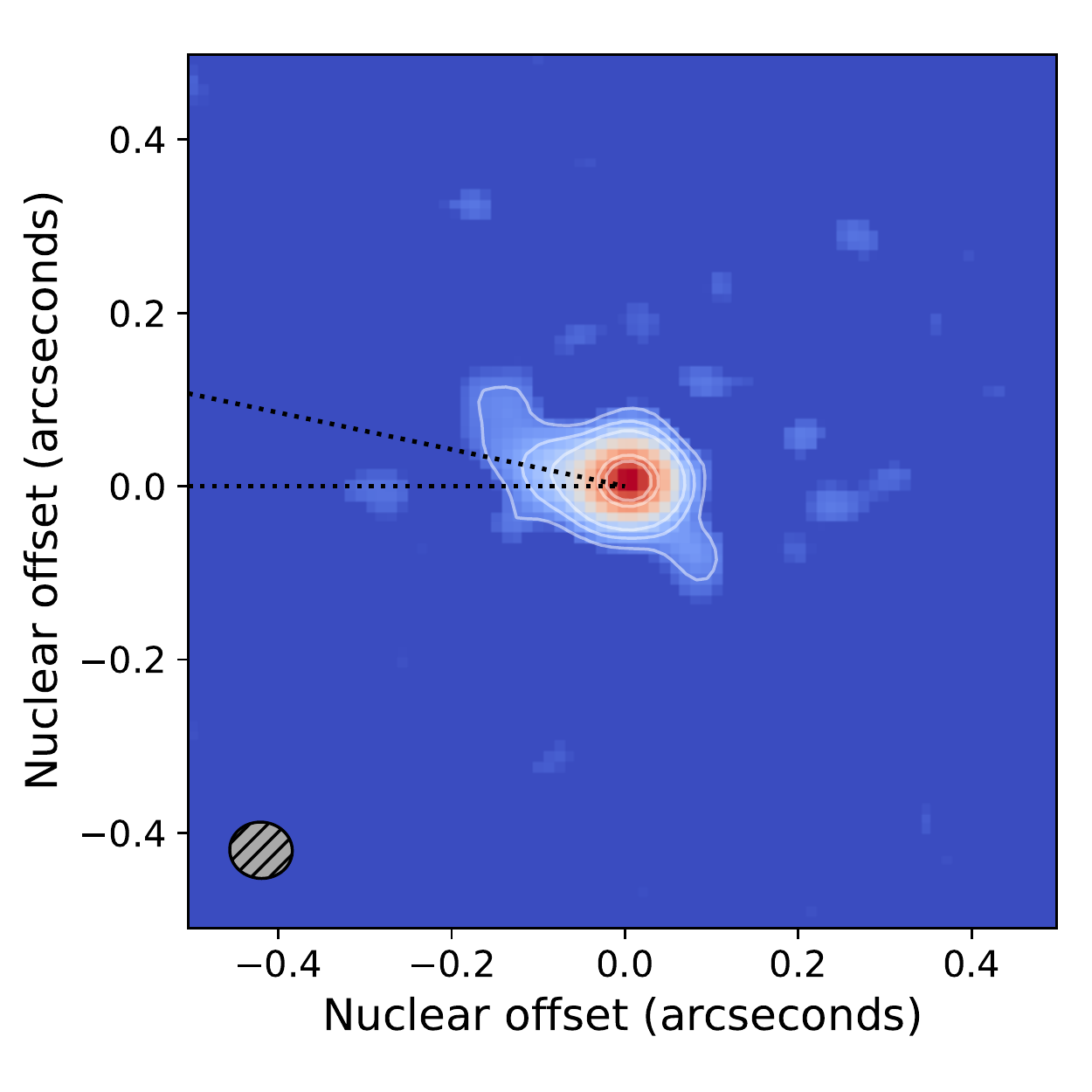}
\vspace{-5mm}
\caption{{ALMA} image of IC 4329A at 100 GHz. \presub{The colour scheme runs from twice the noise level up to the peak value of the core, with square-root scaling. Contours are 4, 10, 20, 80 and 100 times the noise level.} The Gaussian restoring beam (\presub{plotted bottom-left}) is $0.07\arcsec\times0.065\arcsec$ with a major axis position angle of $85\degr$. The \presub{black dotted} lines demonstrate position angles of $78\degr$ and $90\degr$ \edtai{(measured east of north)}.}
\label{fig:alma}
\end{figure}

In Fig. \ref{fig:alma}, we present a high resolution (beam diameter $\sim 0.07\arcsec$, corresponding to $\sim 10$ pc) 100 GHz image of IC 4329A made from a September 2021 observation by the {Atacama Large Millimeter/sub-millimeter Array} ({ALMA}).\footnote{The observations were taken under the programme 2019.1.00580.S (PI: Y. Inoue) with a total integration time of 254 s. We used public pipeline-processed data products from the {ALMA} archive, reduced using \texttt{casa} version 6.2.1.7. Three spectral windows were combined, and the imaging was done using the \texttt{casa} task \texttt{TCLEAN}, adopting a Briggs weighting scheme with a robustness parameter of 0.5. The 12-m array configuration used for the observations allow a maximum recoverable scale of 1.2\arcsec, so the images are not sensitive to any emitting structures larger than this.} \presub{We see that the extended emission is consistent with the $\sim 90\degr$ position angle inferred from longer wavelength images, but also with our measured $\sim 78\degr$ X-ray PA (black dotted lines).} We again see extended emission on only one side, but unlike at 1.5 and 5 GHz, the extension is to the East of the core, not the West.
It therefore seems unlikely that the jet is one sided due to Doppler boosting of the approaching jet and Doppler suppression of the receding jet, in which case the extension would be on the same side for all different scales imaged. It is instead more likely that \referee{the jet bulk Lorentz factor is low enough for Doppler boosting to be unimportant, and} the intrinsic brightness of the jet is not uniform, perhaps with regions of enhanced radio emission corresponding to over-densities that the jet is passing through.

We conclude that the X-ray polarisation is broadly consistent with the jet direction, 
as is observed more conclusively for the black hole X-ray binary Cygnus X-1 \citep{Krawczynski2022}, the neutron star X-ray binary Cygnus X-2 \citep{Farinelli2023} and the AGN NGC 4151 \citep{Gianolli2023}. \presub{Fig. \ref{fig:alma} provides a tentative hint that the small scale jet imaged at 100 GHz aligns with the X-ray PA, with the jet perhaps bending around to $\sim 90\degr$ on the larger scales probed by the VLA images. Even if we take} $90\degr$ as the most likely jet position angle, we find that forcing the X-ray polarisation to be parallel or perpendicular to it gives
\edtai{99} per cent confidence ranges on the PD of respectively
\edtai{$3.0\pm2.9$} per cent and
\edtai{$< 1.2$}
per cent. A radially extended corona (which would be polarised parallel to the jet) is therefore much more compatible with our results than a vertically extended corona (which would be polarised perpendicular to the jet)\referee{; see Fig. \ref{fig:schem}}.

\begin{figure*}
\centering
\includegraphics[width=\columnwidth,trim=2.0cm 1.5cm 3.0cm 0.0cm,clip=true]{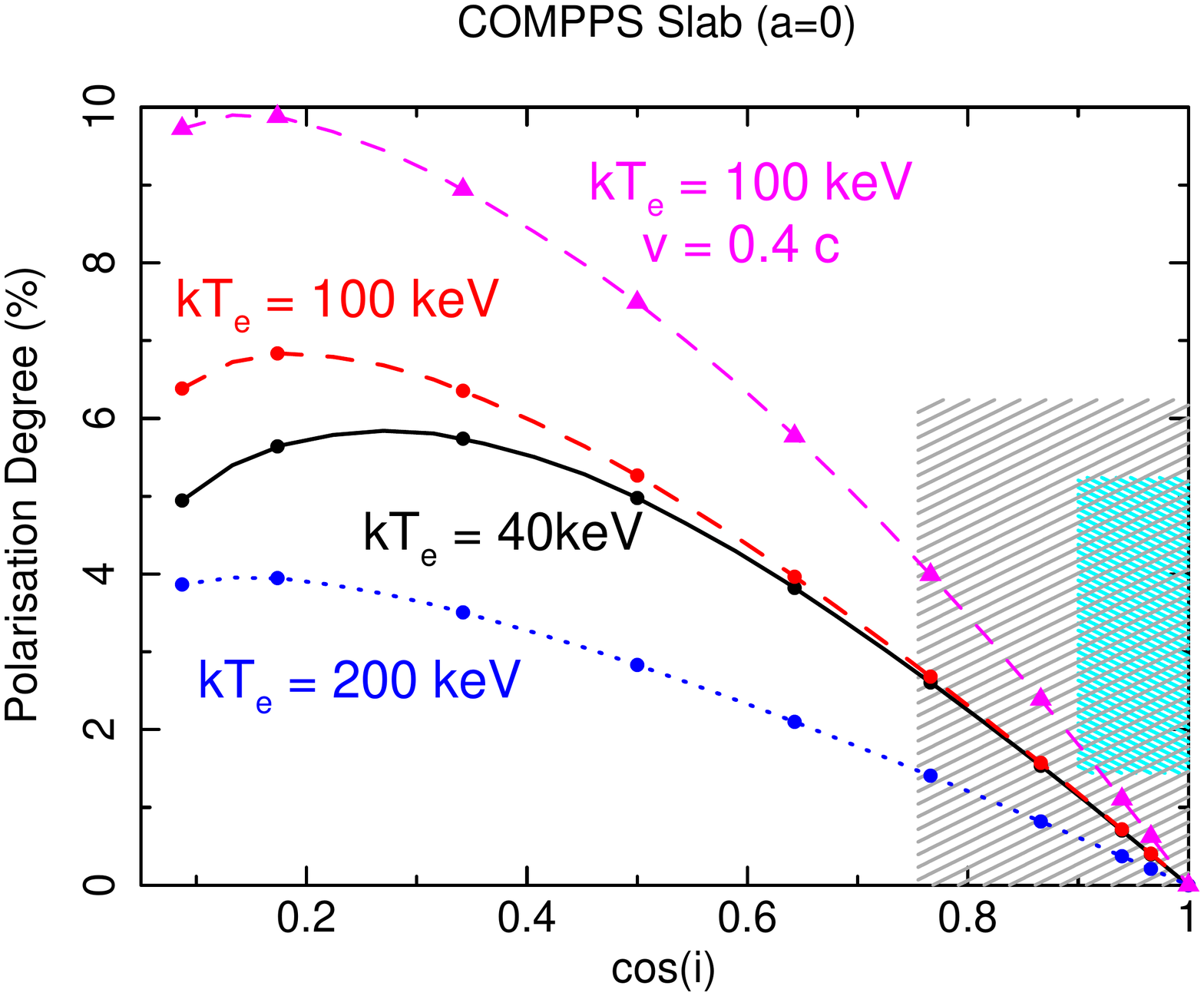}~~
\includegraphics[width=\columnwidth,trim=2.0cm 1.5cm 3.0cm 0.0cm,clip=true]{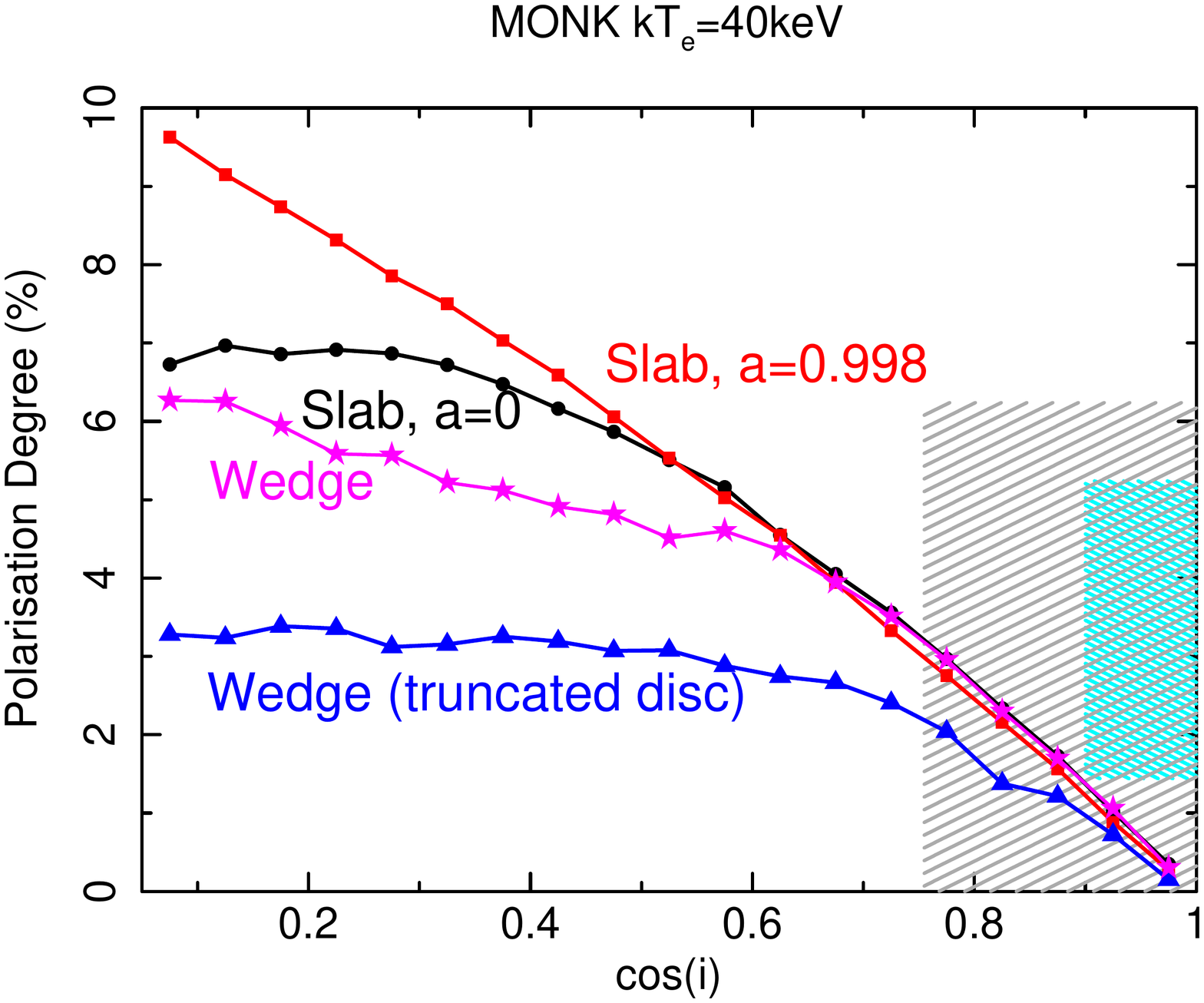}
\vspace{-2mm}
\caption{PD of various coronal geometries as a function of the cosine of the inclination angle. For all geometries considered, the PA is perpendicular to the disc plane. \edtai{The hatched areas represent the observed PD and inclination at $90$ per cent (cyan) and $3 \sigma$ (grey) confidence.} \textit{Left:} A slab corona extending from $6r_\text{g}$ to $100r_\text{g}$, calculated using iterative radiative transfer solver \texttt{COMPPS} for three values of electron temperature. \edtai{Static (circles) and out-flowing (triangles) configurations are considered.} \textit{Right:} Slab and wedge coronal geometries, calculated using the Monte Carlo code \texttt{MONK}, assuming \edtai{a static corona with }an electron temperature of $kT_\text{e}=40$ keV. The slab extends to the ISCO, and two values of black hole spin are considered: $a=0$ (black circles, same case as solid black line in the left panel) and $a=0.998$ (red squares). The wedge has opening angle of $45\degr$ and outer radius $25~r_\text{g}$, and two assumptions are made about the disc: extending to the ISCO (magenta stars), and truncated at $25~r_\text{g}$ (blue triangles).}
\label{fig:monk}
\end{figure*}

\subsection{Spectral fits}

We fit two models to the \ixpe, {\it XMM-Newton} and \textit{NuSTAR} spectra. 
In Model 1, the relativistic and distant reflection components are modelled respectively with \texttt{relxillCp} and \texttt{xillverCp}. Model 2 instead uses \texttt{laor} for the relativistic iron line, a Gaussian for the narrow line, and \texttt{pexriv} for the remainder of the reflection spectrum. Both models prefer a small disc inner radius ($r_{\rm in} \approx 2~r_\text{g}$) and low inclination (at 90 per cent confidence, $i < 26^\circ$ for Model 1 and $i < 45^\circ$ for Model 2), albeit with rather large statistical uncertainties (Fig. \ref{fig:incrin}), and potentially larger systematic uncertainties.
The relativistic component is statistically required for both models. The \texttt{relxillCp} component is required in Model 1 with $5.1\sigma$ significance. \presub{However, it is not possible to disentangle whether it is the broad iron line itself that is required in the Model 1 fit, or whether it is instead the contribution of \texttt{relxillCp} to the broadband continuum. It is therefore encouraging that the \texttt{laor} component is also required in the Model 2 fit ($3.1\sigma$ confidence), providing an indication that a broad line is indeed present in the data}. It is \presub{also somewhat} encouraging that \presub{Models 1 and 2 return consistent estimates of $r_{\rm in}$ and $i$.}

The ionisation parameter of the relativistic reflector and the iron abundance of both reflectors are uncomfortably high in the Model 1 fit. Exploration of parameter space reveals that this combination of parameters is driven by the Compton hump being relatively weak given the strength of the narrow iron line and the relatively hard illuminating spectrum. When we untie the iron abundance of the two reflectors, $A_{\rm Fe}$ of the distant reflector increases further. When we instead replace the distant reflector with a narrow Gaussian iron line, such that only the relativistic reflector contributes a Compton hump, the high $\log\xi$ and $A_{\rm Fe}$ values persist for the relativistic reflector. The high iron abundance therefore appears to be driven by both the relativistic and distant reflectors. We find that an acceptable fit ($\chi^2/{\rm d.o.f.}=1107/1048$) can be achieved by fixing abundance and ionisation to reasonable values ($A_{\rm Fe}=1$ and $\log\xi=3.6$) when we untie the photon index of the illuminating spectrum from the observed value. In this fit, the disc sees a softer coronal spectrum ($\Gamma \approx 2.2$) than we see ($\Gamma \approx 1.7$), which is physically plausible because we are viewing from a low inclination angle whereas the disc is viewing from $i\sim 90^\circ$. The softer illuminating spectrum enables the iron lines to be stronger with respect to the Compton humps without the need for strongly super-solar iron abundance.

A highly super-solar iron abundance is a common feature of reflection fits in the literature \citep{Garcia2018}, and has previously been shown to be at least partially remedied by considering higher disc densities \citep{Tomsick2018}, at least in the case of X-ray binaries. However, here we ignore the $E < 2$ keV energy range that is most sensitive to the density (although we note that reflection models still have some sensitivity to density for $E>3$ keV; \citealt{Ingram2022,Liu2023}) and thus choose to fix it to the canonical value for AGNs of $n_\text{e} = 10^{15}$\,cm$^{-3}$. Moreover, here we assume a single ionisation parameter throughout the disc, whereas in reality we expect the ionisation parameter to be a function of radius \citep{Svoboda2012,Ingram2019}.
\referee{Finally, \cite{Kinch2019} found (albeit for parameters relevant to X-ray binaries), that a very strong iron line can result from self-consistent radiation treatment of general relativistic magnetohyrdodynamic simulations, even when solar abundances are assumed.}

Whereas consistent disc inclination and inner radius values are returned for all the models we explore, we find that different models return different values of the coronal electron temperature. For Model 1, it is $kT_\text{e} \approx 40$
keV, whereas for Model 2, it can be estimated from the exponential cut off as $kT_\text{e} \approx E_\text{cut}/2 \approx 200$ keV. The discrepancy likely comes from the dependence of the shape of the reflection spectrum on $kT_\text{e}$ \citep{Garcia2015a}. We must therefore consider a range of electron temperature values when comparing the observed polarisation to coronal models.

\subsection{Coronal geometry}

The measured PA being roughly aligned with the radio jet favours models in which the corona is radially extended in the plane of the disc, assuming that any misalignment between the jet and the disc rotation axis is small. We can in principle differentiate between different radially extended geometries using the PD \citep{Schnittman2010,Ursini2022}, which we measure to be $p = 3.3 \pm 1.1$ per cent ($1\sigma$ confidence) in the 2--8 keV band. In Fig. \ref{fig:monk}, we plot the predicted PD of various coronal geometries as a function of the cosine of inclination angle. The hatched areas represent the 90 per cent (cyan) and $3\sigma$ (grey) confidence regions of the 2--8 keV PD measured by \ixpe, and the inclination angle returned from our Model 1 spectral fit.  Our low best-fitting inclination (the 90 per cent statistical upper limit is $i < 26\degr$, although there are surely also systematic uncertainties, which are hard to quantify) agrees with previous estimates of the inclination to the inner disc and broad line region \citep{Marin2014,Bentz2023}, and with the Seyfert 1.2 classification of the source in the context of the AGN unification scheme \citep{Urry1995}. Whereas the PD is consistent with zero within $3\sigma$ confidence, at 90 per cent confidence the measured PD is $>1.44$ per cent.

The left panel considers a slab corona located above the disc and extending radially from $6~r_\text{g}$ to $100~r_\text{g}$ (black hole spin assumed to be $a=0$). Here, the polarisation in the source frame is calculated using an iterative radiation transport solver \citep[\texttt{COMPPS},][]{Poutanen1996,Veledina2022}, and then relativistic effects are accounted for using an analytic approximation of the Schwarzschild metric \citep{Poutanen2020}. \edtai{Circular markers represent a static corona, and triangles a vertically out-flowing corona \citep{Poutanen2023}.} Because there is some uncertainty in our spectral fits, we trial three values of the electron temperature: $kT_\text{e} = 40$ keV (as inferred from our Model 1 fit; solid black line), $kT_\text{e} = 100$ keV (dashed lines), and $kT_\text{e} = 200$ keV (as inferred from our Model 2 fit; dotted blue line). For each electron temperature, we choose a coronal optical depth that returns the measured photon index of $\Gamma = 1.66$ (corresponding to the Thomson optical depth of the slab $\tau =2.3$, 0.95 and 0.43 respectively).

We see that PD typically increases with inclination angle, which occurs because the projection of the corona on the sky becomes increasingly asymmetric for larger inclinations.
We also see that the highest values of PD are achieved for the intermediate value of electron temperature. This occurs because of a trade-off between two different effects. First, for a given $\Gamma$, higher temperature corresponds to lower $\tau$ \citep[e.g.][]{Middei2019}. The lower the optical depth, the less likely it is that photons can have multiple scatterings in the vertical direction (i.e. perpendicular to the disc plane), and thus the more likely it is that any photons that did have multiple scatterings were travelling horizontally before they scattered. Because photons are most likely to be scattered \referee{in the plane} perpendicular to their polarisation vector, lower $\tau$ leads to stronger vertical polarisation \referee{(see Fig. \ref{fig:schem})}. The second, competing, effect occurs because higher electron temperature leads to higher electron velocities. This leads to scattered photons being more heavily beamed into the random direction of motion of the electrons, diluting the correlation between the post-scattering trajectory and polarisation of photons \citep{Poutanen1994}.

The right panel of Fig. \ref{fig:monk} presents results from the relativistic Monte Carlo radiative transfer code \texttt{MONK} \citep{Zhang2019}, assuming \edtai{a static corona with temperature} $kT_\text{e} = 40$ keV. The black line with circular markers corresponds to the same geometry as before: a slab extending from $6~r_\text{g}$ to $100~r_\text{g}$ ($a=0$). We see very good agreement with the \texttt{COMPPS} calculation (black solid line in the left hand plot) at the low inclination angles of interest, with some discrepancies at high inclinations that will be explored in future work. We also plot results for a slab extending to the ISCO of a maximally spinning black hole (red squares). We see that relativistic effects significantly effect the polarisation for $i \gtrsim 60\degr$, but not at the lower angles relevant to IC 4329A. We finally consider a uniform density wedge corona with opening angle $45\degr$ extending from the ISCO of a maximally spinning black hole to $25~r_\text{g}$ (Tagliacozzo et al., in prep). In one case, the disc extends within the corona down to the ISCO (magenta stars), and in the other case it is truncated at the outer radius of the corona (blue triangles). These two geometries return a lower PD than the slab because they are less asymmetric.

It is interesting that, in the 90 per cent confidence limit, the observed PD (cyan hatched markings) lies above \edtai{the predictions of all static corona models}
for the inferred inclination angle. \edtai{Assuming instead that the coronal electrons are out-flowing with a mildly relativistic bulk velocity away from the disc plane ($0.4c$ in Fig. \ref{fig:monk}, left) boosts the predicted polarisation into the observed range (due to relativistic aberration). Such an out-flowing model was recently proposed by \cite{Poutanen2023} to explain the high observed PD from Cygnus X-1, and is also consistent with our results.}
We further note that including an out-flowing bulk velocity solves a key problem of the slab model, in that a static slab should not be able to self-consistently produce the hard ($\Gamma<2$) spectrum that is observed because it is irradiated by too high a flux of soft disc photons \citep{Haardt1993,Stern1995,Poutanen2018}. A bulk velocity solves this problem because it reduces the incident seed photon flux via Doppler de-boosting \citep{Beloborodov1999ApJL}. Including a bulk velocity will also enable the wedge and truncated disc geometries to reproduce our polarisation results.

\section{Conclusions}
\label{sec:conclusions}

We have conducted a spectro-polarimetric analysis of IC 4329A using \ixpe, {\it XMM-Newton} and \textit{NuSTAR}. Our \ixpe measurement of 2--8 keV polarisation falls just short of the $3\sigma$ detection threshold (significance $2.97\sigma$). The
\edtai{$1 \sigma$}
confidence limits on PD and PA are
\edtai{$p=3.3\pm1.1$ per cent and $\psi=78\degr \pm 10\degr$,}
respectively. The PA is roughly consistent with the radio jet position angle, once uncertainty on the jet orientation is taken into account. We jointly model the {\it XMM-Newton}, \textit{NuSTAR} and \ixpe spectrum, confirming the presence of a relativistic reflector. Reflection modelling prefers a low inclination ($<26\degr$ with 90 per cent confidence), which is consistent with previous measurements of the viewing angle of the broad line region and the object's Seyfert 1.2 classification within the AGN unification scheme. This constraint coupled with our 90 per cent confidence lower limit on PD of 1.4 per cent tentatively favours more asymmetric, possibly out-flowing, coronal geometries extended in the disc plane that would produce highly polarised emission aligned with the jet, but the coronal geometry is unconstrained at the $3\sigma$ level. Reflection modelling also favours a small disc inner radius, but does not constrain black hole spin because disc truncation outside of the ISCO is statistically acceptable within the $3\sigma$ limit. Spectro-polarimetric modelling is consistent with the 2--8 keV polarisation being dominated by emission observed directly from the corona, but the contribution from reflection is unconstrained.

\section*{Acknowledgements}

The Imaging X-ray Polarimetry Explorer (\ixpe) is a joint US and Italian mission. The US contribution is supported by the National Aeronautics and Space Administration (NASA) and led and managed by its Marshall Space Flight Center (MSFC), with industry partner Ball Aerospace (contract NNM15AA18C). The Italian contribution is supported by the Italian Space Agency (Agenzia Spaziale Italiana, ASI) through contract ASI-OHBI-2017-12-I.0, agreements ASI-INAF-2017-12-H0 and ASI-INFN-2017.13-H0, and its Space Science Data Center (SSDC) with agreements ASI- INAF-2022-14-HH.0 and ASI-INFN 2021-43-HH.0, and by the Istituto Nazionale di Astrofisica (INAF) and the Istituto Nazionale di Fisica Nucleare (INFN) in Italy. This research used data products provided by the \ixpe Team (MSFC, SSDC, INAF, and INFN) and distributed with additional software tools by the High-Energy Astrophysics Science Archive Research Center (HEASARC), at NASA Goddard Space Flight Center (GSFC). We thank the {\it XMM-Newton} and \textit{NuSTAR} SOCs for granting and performing the respective observations of the source. Part of the French contribution is supported by the Scientific Research National Center (CNRS) and the French Space Agency (CNES). A.I. and M.E. acknowledge support from the Royal Society. We thank the anonymous referee for insightful comments that improved the clarity of the paper.

\section*{Data Availability}

The \ixpe data used in this paper are publicly available in the HEASARC database (\url{https://heasarc.gsfc.nasa.gov/docs/ixpe/archive/}). The {\it XMM-Newton} and \textit{NuSTAR} data underlying this article are subject to an embargo of 12 months from the date of the observations. Once the embargo expires the data will be publicly  available from the {\it XMM-Newton} science archive (\url{http://nxsa.esac.esa.int/}) and the \textit{NuSTAR} archive (\url{https://heasarc.gsfc.nasa.gov/docs/ nustar/nustar_archive.html}). The simulation data supporting the findings of the article will be shared on reasonable request.



\bibliographystyle{mnras}
\bibliography{biblio} 

\vspace{1cm}

\noindent \textbf{Affiliations:} \\
\noindent
\textit{
     $^{1}$School of Mathematics, Statistics, and Physics, Newcastle University, Newcastle upon Tyne NE1 7RU, UK \\
     $^{2}$ASI - Agenzia Spaziale Italiana, Via del Politecnico snc, 00133 Roma, Italy \\
     $^{3}$Dipartimento di Matematica e Fisica, Università degli Studi Roma Tre, Via della Vasca Navale 84, 00146 Roma, Italy \\
     $^{4}$Department of Physics and Astronomy, 20014 University of Turku, Finland \\
     $^{5}$Nordita, KTH Royal Institute of Technology and Stockholm University, Hannes Alfvéns väg 12, SE-10691 Stockholm, Sweden \\
     $^{6}$INAF Istituto di Astrofisica e Planetologia Spaziali, Via del Fosso del Cavaliere 100, I-00133 Roma, Italy\\
     $^{7}$Dipartimento di Fisica, Universit\`a degli Studi di Roma “La Sapienza”, Piazzale Aldo Moro 5, I-00185 Roma, Italy\\
     $^{8}$Dipartimento di Fisica, Università degli Studi di Roma “Tor Vergata”, Via della Ricerca    Scientifica 1, I-00133 Roma, Italy\\
     $^{9}$Université de Strasbourg, CNRS, Observatoire Astronomique de Strasbourg, UMR 7550, 67000 Strasbourg, France \\
     $^{10}$MIT Kavli Institute for Astrophysics and Space Research, Massachusetts Institute of Technology, 77 Massachusetts Avenue, Cambridge, MA 02139, USA \\
     $^{11}$Université Grenoble Alpes, CNRS, IPAG, 38000 Grenoble, France \\
     $^{12}$Department of Physics and Kavli Institute for Particle Astrophysics and Cosmology, Stanford University, Stanford, California 94305, USA \\
     $^{13}$Astronomical Institute of the Czech Academy of Sciences, Bo{\v c}ní II 1401/1, 14100 Praha 4, Czech Republic \\
     $^{14}$Physics Department and McDonnell Center for the Space Sciences, Washington University in St. Louis, St. Louis, MO 63130, USA \\
     $^{15}$Space Science Data Center, Agenzia Spaziale Italiana, Via del Politecnico snc, 00133 Roma, Italy \\
     $^{16}$Astronomical Institute, Charles University, V Hole{\v s}ovi{\v c}kách 2, CZ-18000 Prague, Czech Republic \\
     $^{17}$Istituto Nazionale di Fisica Nucleare, Sezione di Roma "Tor Vergata", Via della Ricerca Scientifica 1, 00133 Roma, Italy \\
     $^{18}$Department of Astronomy, University of Maryland, College Park, Maryland 20742, USA \\
     $^{19}$NASA Marshall Space Flight Center, Huntsville, AL 35812, USA \\
     $^{20}$Istituto Nazionale di Fisica Nucleare, Sezione di Torino, Via Pietro Giuria 1, 10125 Torino, Italy \\
     $^{21}$Dipartimento di Fisica, Università degli Studi di Torino, Via Pietro Giuria 1, 10125 Torino, Italy \\
     $^{22}$Instituto de Astrofísica de Andalucía—CSIC, Glorieta de la Astronomía s/n, 18008 Granada, Spain \\
     $^{23}$INAF Osservatorio Astronomico di Roma, Via Frascati 33, 00078 Monte Porzio Catone (RM), Italy \\
     $^{24}$INAF Osservatorio Astronomico di Cagliari, Via della Scienza 5, 09047 Selargius (CA), Italy \\
     $^{25}$Istituto Nazionale di Fisica Nucleare, Sezione di Pisa, Largo B. Pontecorvo 3, 56127 Pisa, Italy \\
     $^{26}$Dipartimento di Fisica, Università di Pisa, Largo B. Pontecorvo 3, 56127 Pisa, Italy \\
     $^{27}$INAF Osservatorio Astrofisico di Arcetri, Largo Enrico Fermi 5, 50125 Firenze, Italy \\
     $^{28}$Dipartimento di Fisica e Astronomia, Università degli Studi di Firenze, Via Sansone 1, 50019 Sesto Fiorentino (FI), Italy \\
     $^{29}$Istituto Nazionale di Fisica Nucleare, Sezione di Firenze, Via Sansone 1, 50019 Sesto Fiorentino (FI), Italy \\
     $^{30}$Science and Technology Institute, Universities Space Research Association, Huntsville, AL 35805, USA \\
     $^{31}$Institut für Astronomie und Astrophysik, Universität Tübingen, Sand 1, 72076 Tübingen, Germany \\
     $^{32}$ RIKEN Cluster for Pioneering Research, 2-1 Hirosawa, Wako, Saitama 351-0198, Japan\\
     $^{33}$California Institute of Technology, Pasadena, CA 91125, USA\\
     $^{34}$Yamagata University, 1-4-12 Kojirakawamachi, Yamagatashi 990-8560, Japan \\
     $^{35}$University of British Columbia, Vancouver, BC V6T 1Z4, Canada \\
     $^{36}$International Center for Hadron Astrophysics, Chiba University, Chiba 263-8522, Japan \\
     $^{37}$Institute for Astrophysical Research, Boston University, 725 Commonwealth Avenue, Boston, MA 02215, USA \\
     $^{38}$Department of Astrophysics, St. Petersburg State University, Uni- versitetsky pr. 28, Petrodvoretz, 198504 St. Petersburg, Russia \\
     $^{39}$Department of Physics and Astronomy and Space Science Center, University of New Hampshire, Durham, NH 03824, USA \\
     $^{40}$Finnish Centre for Astronomy with ESO, 20014 University of Turku, Finland \\
     $^{41}$Istituto Nazionale di Fisica Nucleare, Sezione di Napoli, Strada Comunale Cinthia, 80126 Napoli, Italy \\
     $^{42}$Graduate School of Science, Division of Particle and Astrophysical Science, Nagoya University, Furocho, Chikusaku, Nagoya, Aichi 464-8602, Japan \\
     $^{43}$Hiroshima Astrophysical Science Center, Hiroshima University, 1- 3-1 Kagamiyama, Higashi-Hiroshima, Hiroshima 739-8526, Japan \\
     $^{44}$University of Maryland, Baltimore County, Baltimore, MD 21250, USA \\
     $^{45}$NASA Goddard Space Flight Center, Greenbelt, MD 20771, USA \\
     $^{46}$Center for Research and Exploration in Space Science and Technology, NASA/GSFC, Greenbelt, MD 20771, USA \\
     $^{47}$Department of Physics, The University of Hong Kong, Pokfulam, Hong Kong \\ 
     $^{48}$Department of Astronomy and Astrophysics, Pennsylvania State University, University Park, PA 16802, USA \\
     $^{49}$Center for Astrophysics | Harvard \& Smithsonian, 60 Garden St, Cambridge, MA 02138, USA \\
     $^{50}$INAF Osservatorio Astronomico di Brera, Via E. Bianchi 46, 23807 Merate (LC), Italy \\
     $^{51}$Dipartimento di Fisica e Astronomia, Università degli Studi di Padova, Via Marzolo 8, 35131 Padova, Italy \\
     $^{52}$Mullard Space Science Laboratory, University College London, Holmbury St Mary, Dorking, Surrey RH5 6NT, UK \\
     $^{53}$Anton Pannekoek Institute for Astronomy \& GRAPPA, University of Amsterdam, Science Park 904, 1098 XH Amsterdam, The Netherlands \\
     $^{54}$Guangxi Key Laboratory for Relativistic Astrophysics, School of Physical Science and Technology, Guangxi University , Nanning 530004, China
}






\bsp	
\label{lastpage}
\end{document}